# Survey the storage systems used in HPC and BDA ecosystems.

Priyam Shah
Computer Department
(Student)
Illinois Institute of Technology
*(Student)*
Chicago, USA
pshah129@hawk.iit.edu

Jie Ye
Computer Department
*(Student)*
Illinois Institue of Technology
*(Student)*
Chicago, USA
jye20@hawk.iit.edu

Xian-He Sun
Computer Department
*(Professor)*
Illinois Institute of Technology
*(Professor)*
Chicago, USA
sun@iit.edu

***Abstract*— The advancement in HPC and BDA ecosystem demands a better understanding of the storage systems to plan effective solutions. The amount of data being generated from the ever-growing devices over years have increased tremendously. To make applications access data more efficiently for computation, HPC and BDA ecosystems adopt different storage systems. Each storage system has its pros and cons. Therefore, it is worthwhile and interesting to explore the storage systems used in HPC and BDA respectively. Also, it's inquisitive to understand how such storage systems can handle data consistency and fault tolerance at a massive scale. In this paper, we're surveying four storage systems: Lustre, Ceph, HDFS, and CockroachDB. Lustre and HDFS are some of the most prominent file systems in HPC and BDA ecosystem. Ceph is an upcoming filesystem and is being used by supercomputers. CockroachDB is based on NewSQL systems a technique that is being used in the industry for BDA applications. The study helps us to understand the underlying architecture of these storage systems and the building blocks used to create them. The protocols and mechanisms used for data storage, data access, data consistency, fault tolerance, and recovery from failover are also overviewed. The comparative study will help system designers to understand the key features and architectural goals of these storage systems to select better storage system solutions.**

## 1. Introduction

Storage Systems are an integral part of any architecture. In general, they can be defined as a combination of storage devices (hardware) and file systems (software). The advancement in HPC and BDA ecosystem demands a better understanding of storage systems to plan effective storage solutions. The applications of HPC and BDA may have remained the same but the amount of data being generated from the ever-growing devices over years has increased tremendously. In order to make applications optimize data usage efficiently, HPC and BDA ecosystems adopt different storage systems. Each storage system has its pros and cons. Therefore, It is worthwhile and interesting to explore the storage systems used in HPC and BDA respectively. Also, It's inquisitive to understand how such storage systems can handle data consistency and fault tolerance at a massive scale. In this paper we have surveyed four storage systems Lustre, Ceph, HDFS and CockroachDB. We reviewed Lustre and HDFS because they're the most prominent file systems in HPC and BDA ecosystem. Ceph is an upcoming filesystem for HPC and is being used by supercomputers. CockroachDB is based on NewSQL RDBMS systems a recent technique which is being used in the industry for BDA. To understand the differences between the older and newer filesystem techniques we selected above four storage systems.

Unfortunately, processor, memory, and network technologies are evolving at varying speeds. Clock frequencies do not increase significantly over the years, and even Moore's Law slows down as technology reaches its economic and physical limits [5]. However, due to the heavy use of parallel processing and distributed computing, computing power continues to grow dramatically [6]. The same doesn't apply to storage technology. They have not benefited from comparable advances, so only a small portion of the calculation results can be permanently stored [7]. This discrepancy is sometimes referred to as a memory wall. This requires the user to determine what information is considered worthwhile to store [8]. In addition to storage challenges, politics and practicality demand limiting next-generation exascale systems to 20 MW output [8]. According to, IDC the sum of all digital data, whether created, captured, or duplicated, will increase from 33 zettabytes (ZB) in 2018 to 175 ZB by 2025 [3]. The growing demand for computation and data-intensive applications requires a better understanding of storage systems and their bottlenecks.

### 1.1 Need for Storage Systems in HPC Ecosystem

Supercomputers are a valuable tool for scientific and industrial users. They allow you to perform experiments and gain knowledge in areas that are too expensive, too dangerous, or impossible with other available technologies [4]. Large-scale modeling, simulation, and analysis are used to optimize existing technologies, glimpse the future, and understand phenomena without direct means of imaging or observation. Typical workloads for high-performance computing (HPC) are climate simulation, computational fluid forecasts, and computational fluid dynamics and finite element methods in physics, engineering, and astrophysics [4]. In biology and chemistry, protein folding and molecular dynamics are particularly computationally intensive. With the advent of precision medicine, HPC is becoming more important at the individual level as well. To solve these tasks, many scientific applications are frequently reading and writing large amounts of data to the attached storage systems. The above applications not only solve crucial problems but also contribute to human advancements over years.

### 1.2 Need for Storage Systems in BDA Ecosystem

The data is processed to generate information that can be used later for a variety of purposes. Data mining and knowledge discovery are two areas that we have been actively working on to extract useful information from raw data, make predictions, identify patterns, and create applications that facilitate decision-making [9]. However, with the advent of social media and smart devices, data is no longer a simple dataset that can be processed by traditional tools and technologies [10]. The growing popularity of digitization and the latest technologies such as smartphones and gadgets has contributed significantly to the flood of data. Moreover, this data is not just high in volume, but it also includes data of varied kinds that are generated on a regular basis. The biggest challenge in dealing with this "big data problem" is that current or traditional systems cannot store and process such data. This required a scalable system that could store data in a variety of formats and process them into meaningful analytical solutions [11]. Various technologies are available for this purpose, and organizations can use data stores such as HBase [12], HDFS [13], MongoDB [14], execution engines such as Impala [15] and Spark [16], and R. Programming languages such as [17] and Python [18]. Big data storage [19] is a general term used to describe a storage infrastructure designed to store, manage, and retrieve data. In such an infrastructure, data is stored for ease of usage, access, and processing. In addition, such infrastructure can be expanded according to the requirements of the application or service.

### 1.3 File Systems

Providing reliable, efficient, and easy-to-use file systems is one of the biggest challenges in today's HPC and BDA ecosystems, as various scientific and social applications generate and analyze vast amounts of data. The file system provides an interface to the underlying storage device and links identifiers such as filenames to the corresponding physical addresses in the storage hardware. This allows for more comfortable and simplified use of storage devices. Traditionally, directories and files have been used to implement the concept of hierarchy. In addition to the actual file content, metadata such as file size and access time is also managed. Several filesystems that offer a wide range of features have been proposed and established over the years.

The need for high-throughput simultaneous read and write capabilities in HPC applications has led to the development of parallel and distributed file systems. The data can thus be distributed across a large number of storage devices and combine special properties to increase throughput and system capacity. However, due to the proliferation of data, processing vast amounts of information requires a more sophisticated and professional approach. At the same time, new and more powerful storage and network technologies are being developed that challenge each feature. Few well-known file systems in HPC ecosystem include Lustre, Spectrum Scale, BeeGFS, OrangeFS, Ceph, and GlusterFS.

The main task of the big data storage system is to support the storage of large numbers of files and objects, as well as the input and output operations of the stored data. Architectures typically used to store big data include clusters of network-attached storage and pools of directly attached storage [19]. At the heart of these infrastructures are compute server nodes that support the acquisition and processing of big data. Most of these storage infrastructures support big data storage solutions such as Hadoop [20], NoSQL [21], and NewSQL.

## 2. HPC Storage Systems

### 2.1 Lustre

Lustre is a parallel file system that is used on supercomputers. It is licensed under the GNU General Public License (GPLv2) and can be extended and improved. Because of its high performance, Lustre is used on more than half of the 100 fastest supercomputers in the world. The figure 1 shows the file system's architecture which distinguishes between clients and servers. Clients use RPC messages to communicate with the servers, which perform the actual I/O operations. While all clients are identical, the servers can have different roles: Object Storage Servers (OSS) manage the file system's data in the form of objects; clients can access byte ranges within the objects. Metadata Servers (MDS) manage the file system's metadata; after retrieving the metadata, clients can independently contact the appropriate OSSs. Each server is connected to possibly multiple targets (OSTs/MDTs) that store the actual file data or metadata, respectively. The MGS stores configuration information for all the Lustre file systems in a cluster and provides this information to other Lustre components. Each Lustre target contacts the MGS to provide information, and Lustre clients contact the MGS to retrieve information. It is preferable that the MGS have its own storage space so that it can be managed independently [22].

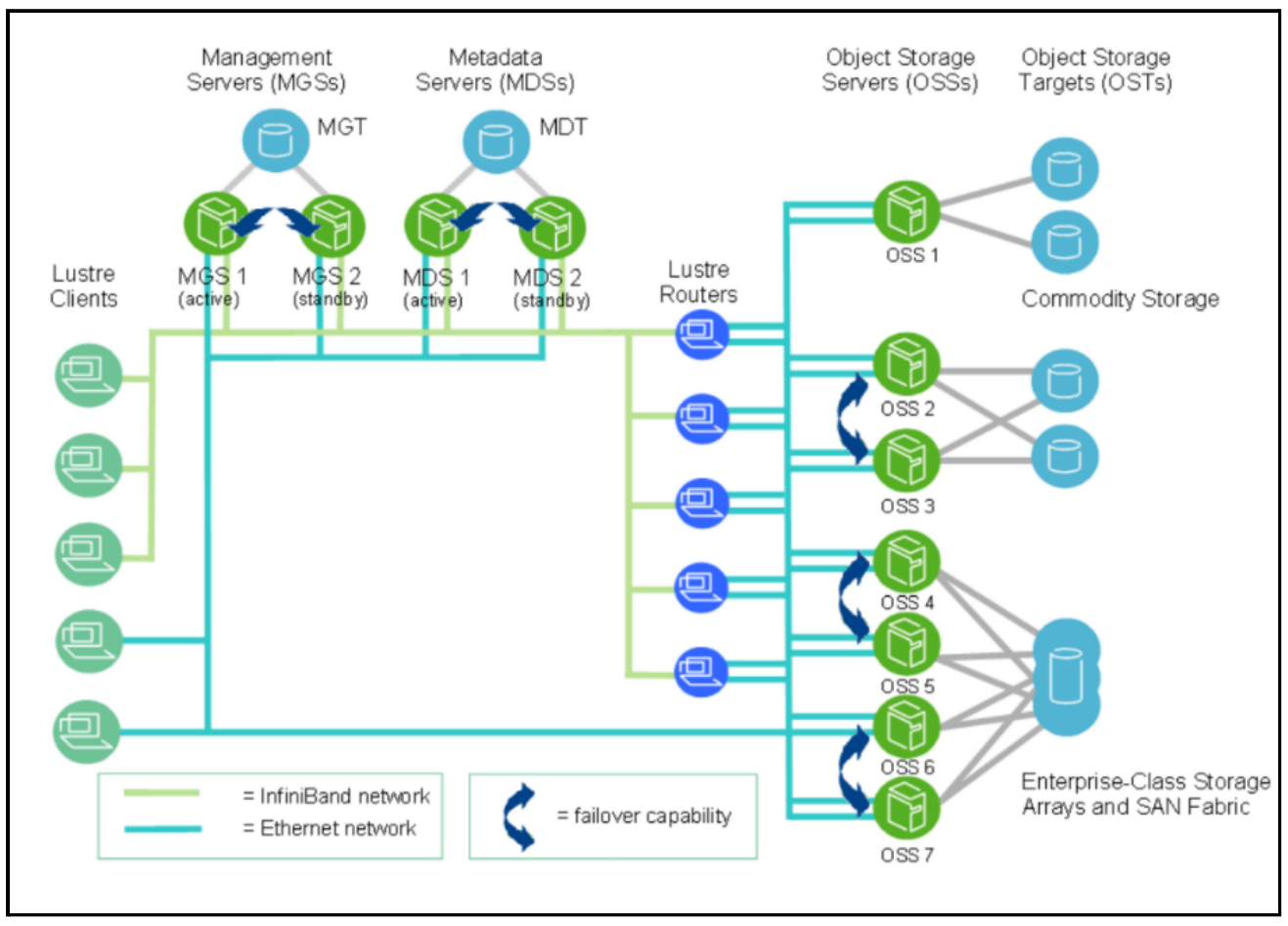

**Fig 1. Lustre Architecture [30]**

The Lustre Networking layer (LNet) operates above the Lustre Network Driver (LND) layer in a manner similar to the way the network layer operates above the data link layer. LNet layer is connectionless, asynchronous, and does not verify that data has been transmitted while the LND layer is connection-oriented and typically does verify data transmission [22].

Lustre runs in kernel space, that is, most functionality has been implemented in the form of kernel modules, which has advantages and disadvantages. On the one hand, by using the kernel's virtual file system (VFS) Lustre can provide a POSIX-compliant file system that is compatible with existing applications. On the other hand, each file system operation requires a system call, which can be expensive when dealing with high-performance network and storage devices [22].

### 2.1.1. Data Storage in Lustre

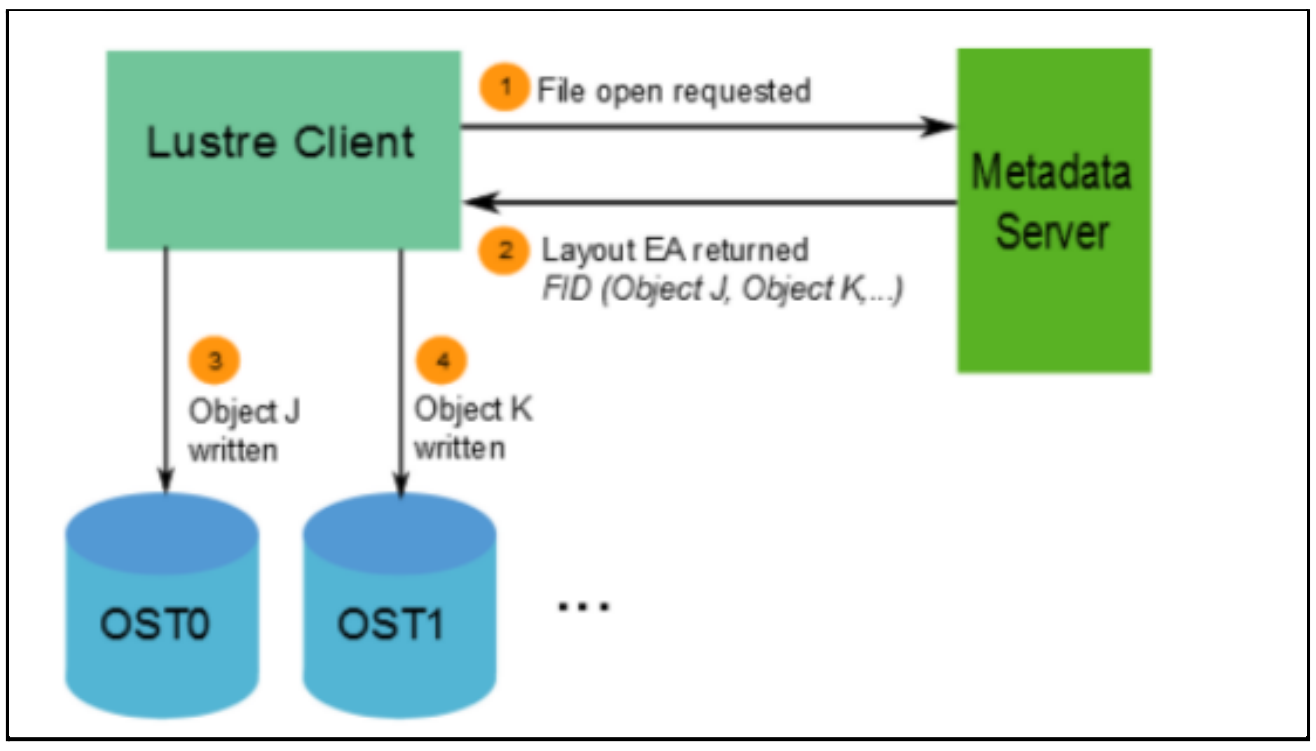

**Fig 2. Lustre client requesting file data to write. [30]**

Lustre File Identifiers (FIDs) are used internally for identifying files or objects, similar to inode numbers in local filesystems. An FID is a 128-bit identifier, which contains a unique 64-bit sequence number (SEQ), a 32-bit object ID (OID), and a 32-bit version number. The sequence number is unique across all Lustre targets in a file system (OSTs and MDTs). This allows multiple MDTs and OSTs to uniquely identify objects without depending on identifiers in the underlying filesystem (e.g. inode numbers) that are likely to be duplicated between targets. The LFSCK file system consistency checking tool provides functionality that helps in verifying invalidity or missing FID. Information about where file data is located on the OST(s) is stored as an extended attribute called layout EA in an MDT object identified by the FID. The above figure 2 explains a simple file write data request made by Lustre client. First the request goes to metadata server which returns and Extended Attribute (EA) of object addresses then respective OSTs are contacted for the operation all these occur over the LNet. One of the main factors leading to the high performance of Lustre file systems is the ability to stripe data across multiple OSTs in a round-robin fashion.

### 2.1.2 Data Consistency in Lustre

Lustre file system has few consistency issues like dangling references, orphan objects, and repeated references. The consistency framework has the following solutions:

- FID-in-LMA (Lustre Metadata Attribute)**:** Lustre object stores its FID in the XATTR_NAME_LMA extended attribute (EA) for related object index mapping consistency and self-verification.
- linkEA: The MDT-object stores its position (in namespace) information (the name and the parent FID) as XATTR_NAME_LINK EA.
- parent FID for OST-object: The OST-object stores the FID of its parent MDT-object that references the OST-object as XATTR_NAME_FID EA.

To verify consistency, it provides Lustre consistency verification tools - LFSCK that can verify the objects in the whole/partial system.

### 2.1.3 Fault-Tolerance in Lustre

In a high-availability (HA) system, unscheduled downtime is minimized by using redundant hardware and software components and software components that automate recovery when a failure occurs. Availability is accomplished by replicating hardware and/or software so that when a primary server fails or is unavailable, a standby server can be switched into its place to run applications and associated resources. This process, called failover, is automatic in an HA system and, in most cases, completely application transparent.

To establish a highly available Lustre file system, power management software or hardware and high availability (HA) software are used to provide the following failover capabilities:

- Resource fencing - Protects physical storage from simultaneous access by two nodes.
- Resource management - Starts and stops the Lustre resources as a part of failover, maintains the cluster state, and carries out other resource management tasks.
- Health monitoring - Verifies the availability of hardware and network resources and responds to health indications provided by the Lustre software.

#### 2.1.3.1 Types of Failover Configurations

- Active/passive pair – Figure 3 shows this configuration, the active node provides resources and serves data, while the passive node is usually standing by idle. If the active node fails, the passive node takes over and becomes active.

- Active/active pair – Figure 4 shows this configuration, both nodes are active, each providing a subset of resources.

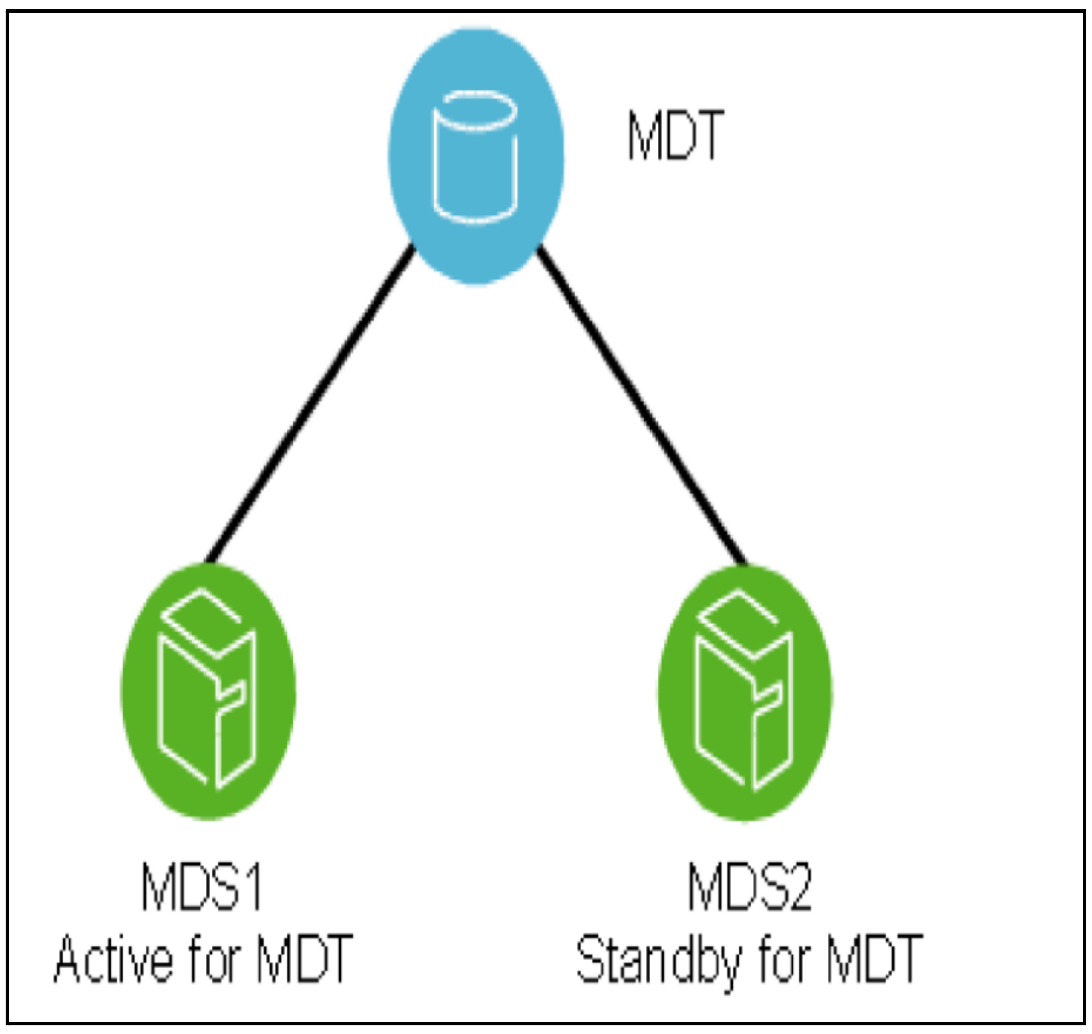

**Fig 3: Lustre failover configuration for active/passive MDT [22]**

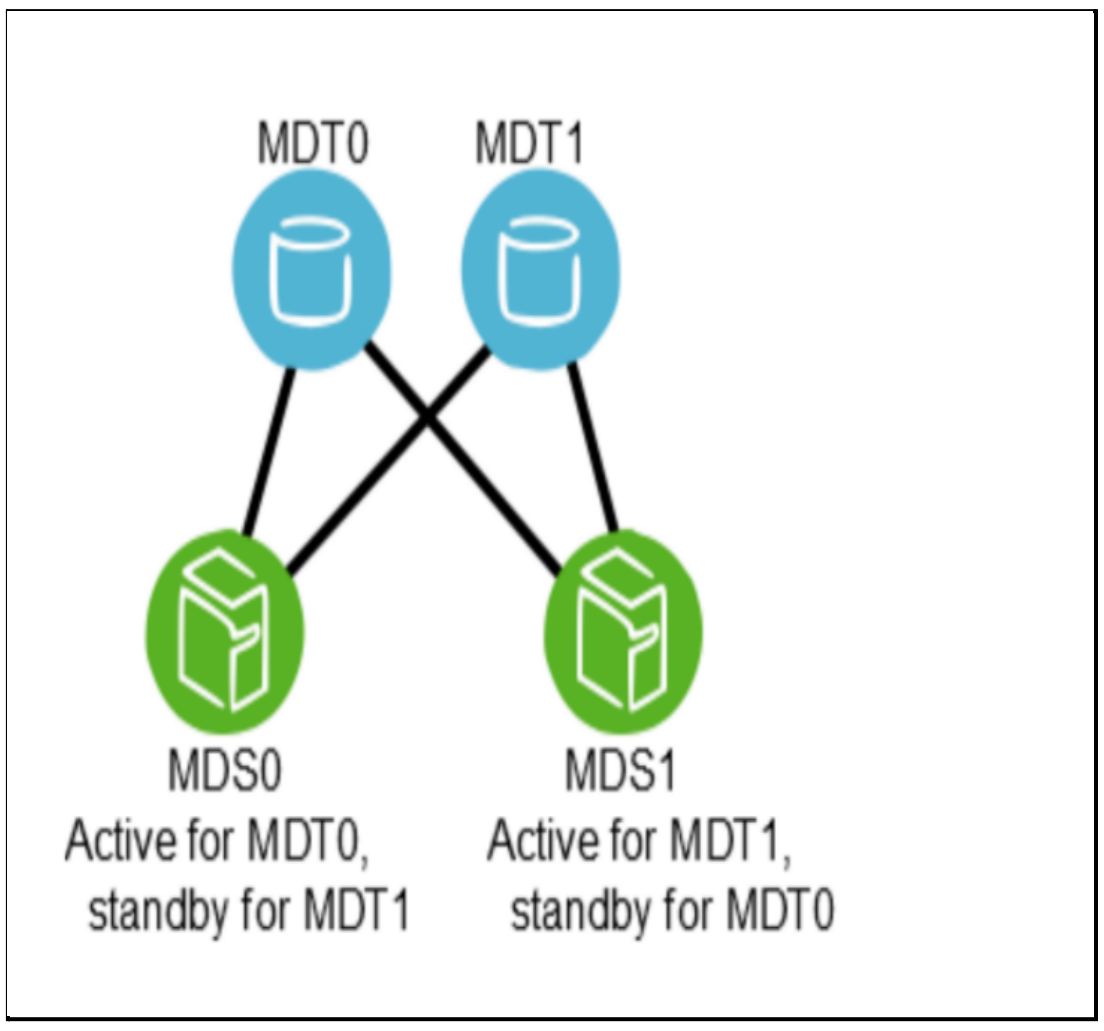

**Fig 4: Lustre failover configuration for active/active MDT [22]**

### 2.1.4 Recovery in Lustre

The recovery feature provided in the Lustre software is responsible for dealing with node or network failure and returning the cluster to a consistent, performant state. Because the Lustre software allows servers to perform asynchronous update operations to the on-disk file system (i.e., the server can reply without waiting for the update to synchronously commit to disk), the clients may have a state in memory that is newer than what the server can recover from disk after a crash.

A handful of different types of failures can cause recovery to occur: Client (compute node) failure, MDS failure (and failover), OST failure (and failover) and Transient network partition.

For Lustre, all Lustre file system failure and recovery operations are based on the concept of connection failure; all imports or exports associated with a given connection are considered to fail if any of them fail. Following are the recovery methods:

a) "Imperative Recovery" feature allows the MGS to actively inform clients when a target restarts after a failure, failover, or other interruption to speed up recovery.
b) "Metadata Replay" feature provides information on recovering from a corrupt file system.
c) "Commit on Share" feature provides information on resolving orphaned objects, a common issue after recovery.

## 2.2 Ceph

Ceph is a free and open-source platform that offers file-, block- and object-based data storing on a single distributed cluster [24]. Figure 5 shows the Ceph architecture. The system implements distributed object storage on a base of the Reliable Autonomic Distributed Object Store (RADOS) system [25]. It is responsible for data migration, replication, failure detection, and failure recovery to the cluster. Integration of the near-POSIX-compliant CephFS file system allows many applications to utilize the benefits and capabilities of the scalable environment. Ceph makes use of intelligent Object Storage Devices (OSDs). These units provide file I/O (reads and writes) for all clients who interact with them. Data and metadata are decoupled because all the operations for metadata altering are performed by Metadata Servers (MDSs). Ceph dynamically distributes the metadata management and responsibility for the file system directory hierarchy among tens or even hundreds of those MDSs.

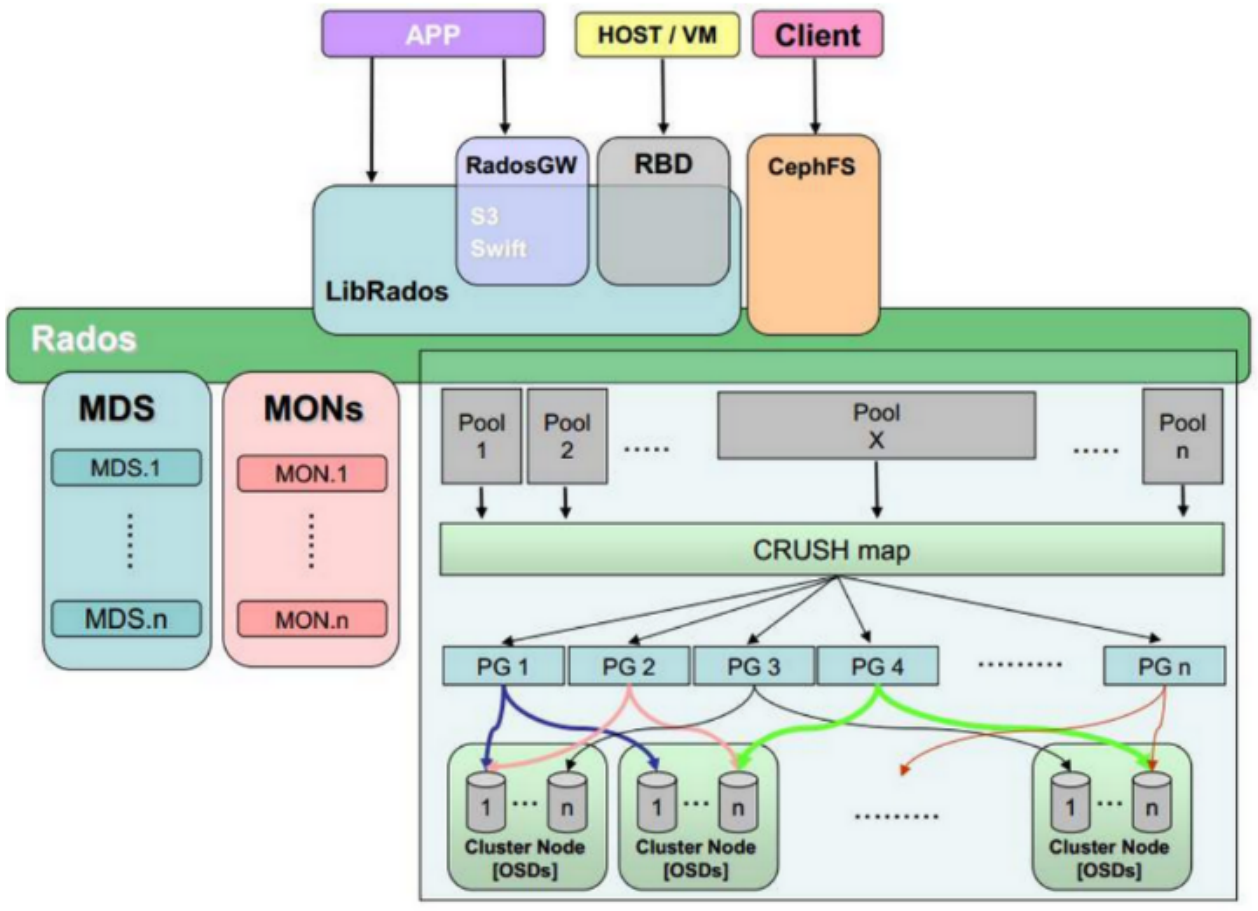

**Fig 5: Ceph Architecture**

A Ceph Storage Cluster consists of multiple types of daemons [28]:

- Ceph Monitor: A Ceph Monitor maintains a master copy of the cluster map. A cluster of Ceph monitors ensures high availability. Storage cluster clients retrieve a copy of the cluster map from the Ceph Monitor.
- Ceph OSD Daemon: A Ceph OSD Daemon checks its own state and the state of other OSDs and reports back to monitors.
- Ceph Manager: A Ceph Manager acts as an endpoint for monitoring, orchestration, and plug-in modules.

- Ceph Metadata Server: A Ceph Metadata Server (MDS) manages file metadata when CephFS is used to provide file services.

**2.2.1 CRUSH Algorithm (Controlled, Scalable, Decentralized Placement of Replicated Data):**
Ceph Clients and Ceph OSD Daemons both use the CRUSH algorithm to efficiently compute information about object location, instead of having to depend on a central lookup table. CRUSH provides a better data management mechanism compared to older approaches and enables massive scale by cleanly distributing the work to all the clients and OSD daemons in the cluster. CRUSH uses intelligent data replication to ensure resiliency, which is better suited to hyper-scale storage [28].

**2.2.2 Data-Storage in Ceph**
The Ceph Storage Cluster receives data from Ceph Clients–whether it comes through a Ceph Block Device, Ceph Object Storage, the Ceph File System, or a custom implementation you create using librados– which is stored as RADOS objects. Each object is stored on an Object Storage Device. Ceph OSD Daemons handle read, write, and replication operations on storage drives. With the older File store back end, each RADOS object was stored as a separate file on a conventional filesystem (usually XFS) [28]. With the new and default BlueStore back end, objects are stored in a monolithic database-like fashion as shown in Figure 6.

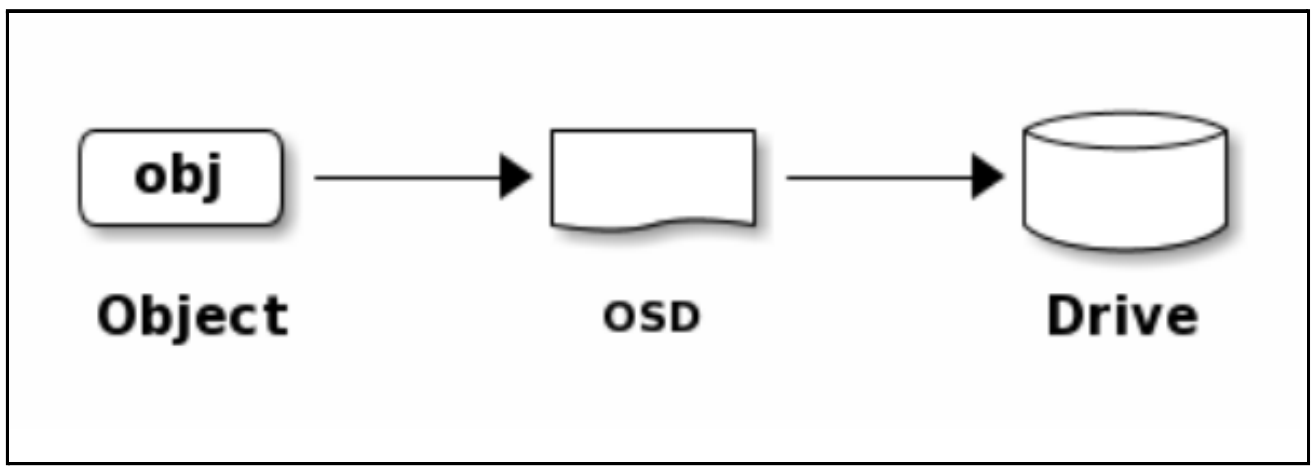

**Fig 6: Ceph data storage high level [28]**

Ceph OSD Daemons store data as objects in a flat namespace (e.g., no hierarchy of directories) as shown in Figure 7. An object has an identifier, binary data, and metadata consisting of a set of name/value pairs. The semantics are completely up to Ceph Clients. For example, CephFS uses metadata to store file attributes such as the file owner, created date, last modified date, and so forth [28].

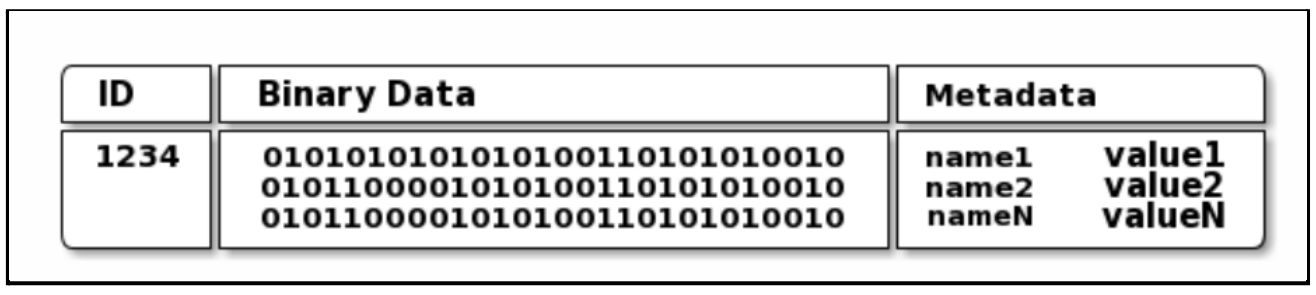

**Fig 7: Ceph namespace for data storage [28]**

The Ceph storage system supports the notion of 'Pools', which are logical partitions for storing objects. Ceph Clients retrieve a Cluster Map from a Ceph Monitor and write objects to pools. The pool's size or the number of replicas, the CRUSH rule, and the number of placement groups determine how Ceph will place the data.

**2.2.3 Data-Access in Ceph**
OSDs Service Clients Directly: Since any network device has a limit to the number of concurrent connections it can support, a centralized system has a low physical limit at high scales. By enabling Ceph Clients to contact Ceph OSD Daemons directly, Ceph increases both performance and total system capacity simultaneously, while removing a single point of failure. Ceph Clients can maintain a session when they need to and with a particular Ceph OSD Daemon instead of a centralized server [28].

**2.2.4 Data-Consistency in Ceph**
Data Scrubbing: As part of maintaining data consistency and cleanliness, Ceph OSD Daemons can scrub objects. That is, Ceph OSD Daemons can compare their local objects metadata with its replicas stored on other OSDs. Scrubbing happens on a per-Placement Group base. Scrubbing (usually performed daily) catches mismatches in size and other metadata. Ceph OSD Daemons also perform deeper scrubbing by comparing data in objects bit-for-bit with their checksums. Deep scrubbing (usually performed weekly) finds bad sectors on a drive that wasn't apparent in a light scrub [28].

**2.2.5 Fault Tolerance in Ceph**
If an MDS daemon stops communicating with the monitor, the monitor will wait mds_beacon_grace seconds (default 15 seconds) before marking the daemon as laggy. If the standby is available, the monitor will immediately replace the laggy daemon. Each file system may specify the number of standby daemons to be considered healthy. This number includes daemons in standby replay waiting for a rank to fail. The pool of standby daemons not in replay count towards any file system count [28].

**2.2.6 Recovery in Ceph**
a) Metadata damage and repair: If a file system has inconsistent or missing metadata, it is considered damaged. You may find out about damage from a health message, or in some unfortunate cases from an assertion in a running MDS daemon. Metadata damage can result either from data loss in the underlying RADOS layer (e.g. multiple disk failures that lose all copies of a PG) or from software bugs. CephFS includes some tools that may be able to recover a damaged file system, but to use them safely requires a solid understanding of CephFS internals [28].

b) Data pool damage (files affected by lost data PGS): If a PG is lost in a data pool, then the file system will continue to operate normally, but some parts of some files will simply be missing (reads will return zeros). Losing a data PG may affect many files. Files are split into many objects, so identifying which files are affected by the loss of particular PGs requires a full scan of overall object IDs that may exist within the size of a file. This type of scan may be useful for identifying which files require restoring from a backup [28].

**3. BDA Storage Systema**

### 3.1 CockroachDB

CockroachDB was designed to create the source-available database for both scalability and consistency. CockroachDB was designed to meet the following goals: Offer industry-leading consistency, even on massively scaled deployments. This means enabling distributed transactions, as well as removing the pain of eventual consistency issues and stale reads. Create an always-on database that accepts reads and writes on all nodes without generating conflicts. Allow flexible deployment in any environment, without tying you to any platform or vendor. Support familiar tools for working with relational data (i.e., SQL). [33]

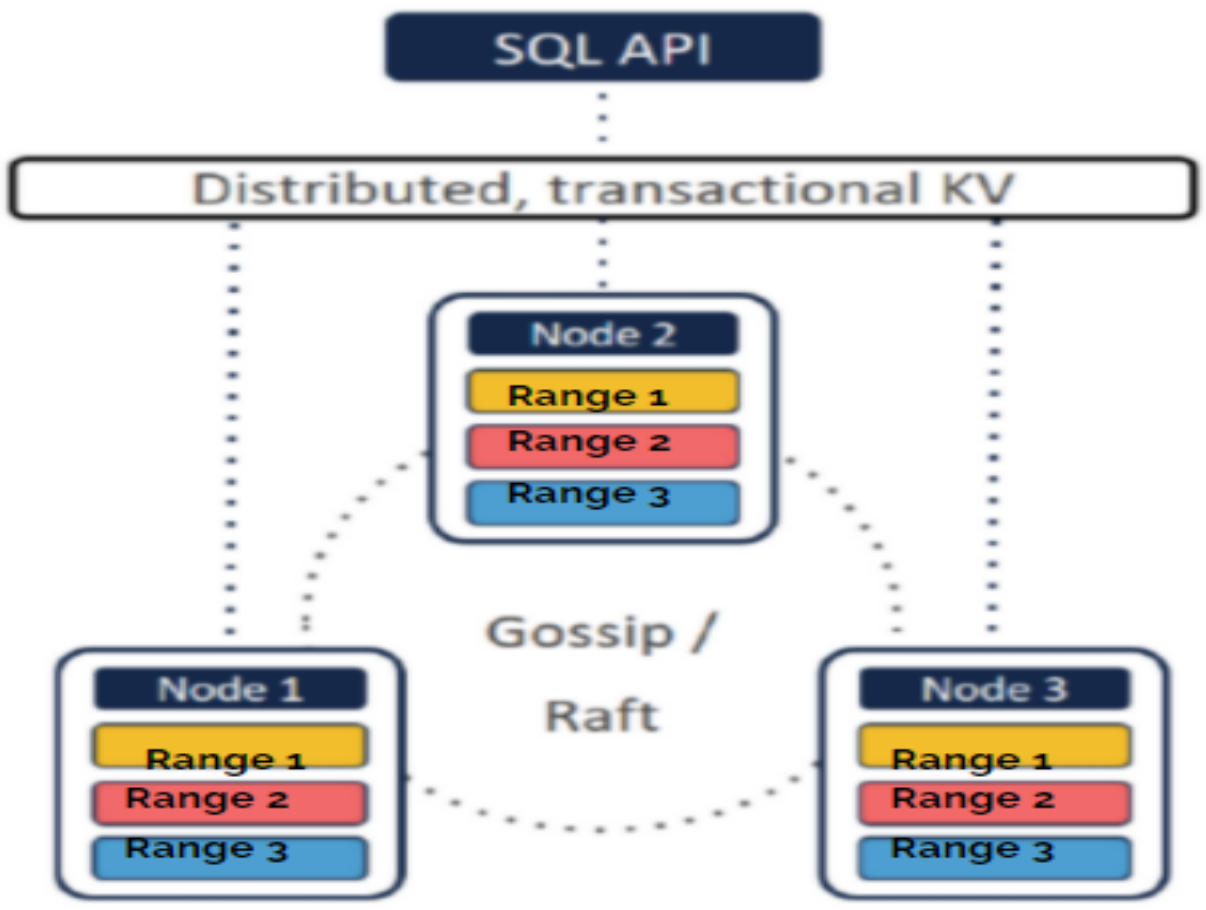

**Fig 8 CockroachDB Architecture [33]**

Figure 8 shows the CockroachDB Architecture, Once the CockroachDB cluster is initialized, developers interact with CockroachDB through a PostgreSQL-compatible SQL API. Since there is symmetrical behavior of all nodes in a cluster, you can send SQL requests to any node; this makes CockroachDB easy to integrate with load balancers. After receiving SQL remote procedure calls (RPCs), nodes convert them into key-value (KV) operations that work with the distributed, transactional key-value stores.[33]

As the RPCs start filling your cluster with data, CockroachDB starts algorithmically distributing your data among the nodes of the cluster, breaking the data up into 512 MiB chunks (ranges). Each range is replicated to at least 3 nodes by default to ensure survivability. This ensures that if any nodes go down, you still have copies of the data which can be used for continuing to serve reads and writes and consistently replicate the data to other nodes. If a node receives a read or write request it cannot directly serve, it finds the node that can handle the request, and communicates with that node. This means you do not need to know where in the cluster a specific portion of your data is stored; CockroachDB tracks it for you and enables symmetric read/write behavior from each node.[33]

CockroachDB's architecture is manifested as a number of layers, each of which interacts with the layers directly above and below it as relatively opaque services. Layers and their purpose in CockroachDB are shown in the table below.

| Layer | Order | Purpose |
|---|---|---|
| **SQL** | 1 | Translate client SQL queries to KV operations. |
| **Transactional** | 2 | Allow atomic changes to multiple KV entries. |
| **Distribution** | 3 | Present replicated KV ranges as a single entity. |
| **Replication** | 4 | Consistently and synchronously replicate KV ranges across many nodes. This layer also enables consistent reads using a consensus algorithm. |
| **Storage** | 5 | Read and write KV data on disk. |

**Table 1: Layers in CockroachDB [33]**

#### 3.1.1 Data-Storage in CockroachDB

The storage layer of CockroachDB's architecture reads and writes data to disk. Each CockroachDB node contains at least one store, specified when the node starts, which is where the cockroach process reads and writes its data on disk. This data is stored as key-value pairs on a disk using the storage engine, which is treated primarily as a black-box API. CockroachDB uses the Pebble storage engine. Pebble is intended to be bi-directionally compatible with the RocksDB on-disk format but differs in that it is written in Go and implements a subset of RocksDB's large feature set. It contains optimizations that benefit CockroachDB. Internally, each store contains two instances of the storage engine one for storing temporary distributed SQL data and one for all other data on the node. In addition, there is also a block cache shared amongst all of the stores in a node. These stores in turn have a collection of range replicas. More than one replica for a range will never be placed on the same store or even the same node. [33]

#### 3.1.2 Data Access in CockroachDB

To make all data in your cluster accessible from any node, CockroachDB stores data in a monolithic sorted map of key-value pairs. This key-space describes all of the data in your cluster, as well as its location, and is divided into what we call "ranges", contiguous chunks of the key-space so that every key can always be found in a single range.

CockroachDB implements a sorted map to enable:

a) Simple lookups: To identify which nodes are responsible for certain portions of the data, queries are able to quickly locate where to find the data they want.
b) Efficient scans: By defining the order of data, it's easy to find data within a particular range during a scan.

The monolithic sorted map is comprised of two fundamental elements: System data, which include meta ranges that describe the locations of data in your cluster (among many other cluster-wide and local data elements), and User data, which store your cluster's table data. [33]

### 3.1.3 Data-Consistency in CockroachDB

To provide consistency, CockroachDB implements full support for ACID transaction semantics in the transaction layer. However, it's important to realize that all statements are handled as transactions, including single statements—this is sometimes referred to as "auto-commit mode" because it behaves as if every statement is followed by a COMMIT. Because CockroachDB enables transactions that can span your entire cluster (including cross-range and cross-table transactions), it achieves correctness using a distributed, atomic commit protocol called Parallel Commits.

Any changes made to the data in a range rely on a consensus algorithm to ensure that the majority of the range's replicas agree to commit the change. This is how CockroachDB achieves the industry-leading isolation guarantees that allow it to provide your application with consistent reads and writes, regardless of which node you communicate with.

CockroachDB relies heavily on multi-version concurrency control (MVCC) to process concurrent requests and guarantee consistency. Much of this work is done by using hybrid logical clock (HLC) timestamps to differentiate between versions of data, track commit timestamps, and identify a value's garbage collection expiration. All of this MVCC data is then stored in Pebble. CockroachDB maintains a timestamp cache, which stores the timestamp of the last time that the key was read. If a write operation occurs at a lower timestamp than the largest value in the read timestamp cache, it signifies there’s a potential anomaly and the transaction must be restarted at a later timestamp. [33]

#### 3.1.3.1 Parallel Commits in CockroachDB

The Parallel Commits feature introduces a new, optimized atomic commit protocol that cuts the commit latency of a transaction in half, from two rounds of consensus down to one. Combined with Transaction pipelining, this brings the latency incurred by common OLTP transactions to near the theoretical minimum: the sum of all read latencies plus one round of consensus latency.

Under the new atomic commit protocol, the transaction coordinator can return to the client eagerly when it knows that the writes in the transaction have succeeded. Once this occurs, the transaction coordinator can set the transaction record's state to COMMITTED and resolve the transaction's write intentions asynchronously. The transaction coordinator is able to do this while maintaining correctness guarantees because it populates the transaction record with enough information (via a new STAGING state, and an array of in-flight writes) for other transactions to determine whether all writes in the transaction are present, and thus prove whether or not the transaction is committed. [33]

### 3.1.4 Fault Tolerance in CockroachDB

Ensuring consistency with nodes offline, though, is a challenge many databases fail. To solve this problem, CockroachDB uses a consensus algorithm to require that a quorum of replicas agrees on any changes to a range before those changes are committed. Because 3 is the smallest number that can achieve quorum (i.e., 2 out of 3), CockroachDB's high availability (known as multi-active availability) requires 3 nodes.

The number of failures that can be tolerated is equal to (Replication factor - 1)/2. For example, with 3x replication, one failure can be tolerated; with 5x replication, two failures, and so on. You can control the replication factor at the cluster, database, and table-level using replication zones.

When failures happen, though, CockroachDB automatically realizes nodes have stopped responding and works to redistribute your data to continue maximizing survivability. This process also works the other way around: when new nodes join your cluster, data automatically rebalances onto it, ensuring your load is evenly distributed.[33]

#### 3.1.4.1 Raft Protocol in CockroachDB

Raft is a consensus protocol—an algorithm that makes sure that your data is safely stored on multiple machines, and that those machines agree on the current state even if some of them are temporarily disconnected.

Raft organizes all nodes that contain a replica of a range into a group--unsurprisingly called a Raft group. Each replica in a Raft group is either a "leader" or a "follower". The leader, which is elected by Raft and long-lived, coordinates all writes to the Raft group. It heartbeats followers periodically and keeps their logs replicated. In the absence of heartbeats, followers become candidates after randomized election timeouts and proceed to hold new leader elections.

### 3.1.5 Recovery in CockroachDB

Each replica can be "snapshotted", which copies all its data as of a specific timestamp. This snapshot can be sent to other nodes during a rebalance when a new node is added. After loading the snapshot, the node gets up to date by replaying all actions from the Raft group's log that have occurred since the snapshot was taken.

## 3.2 Hadoop Distributed File System (HDFS)

The Hadoop Distributed File System (HDFS) is a distributed file system designed to run on commodity hardware. It has many similarities with existing distributed file systems. HDFS is highly fault-tolerant and is designed to be deployed on low-cost hardware. HDFS provides high throughput access to application data and is suitable for applications that have large data sets. HDFS relaxes a few POSIX requirements to enable streaming access to file system data. HDFS was originally built as infrastructure for the Apache Nutch web search engine project. HDFS is now an Apache Hadoop subproject.[34]

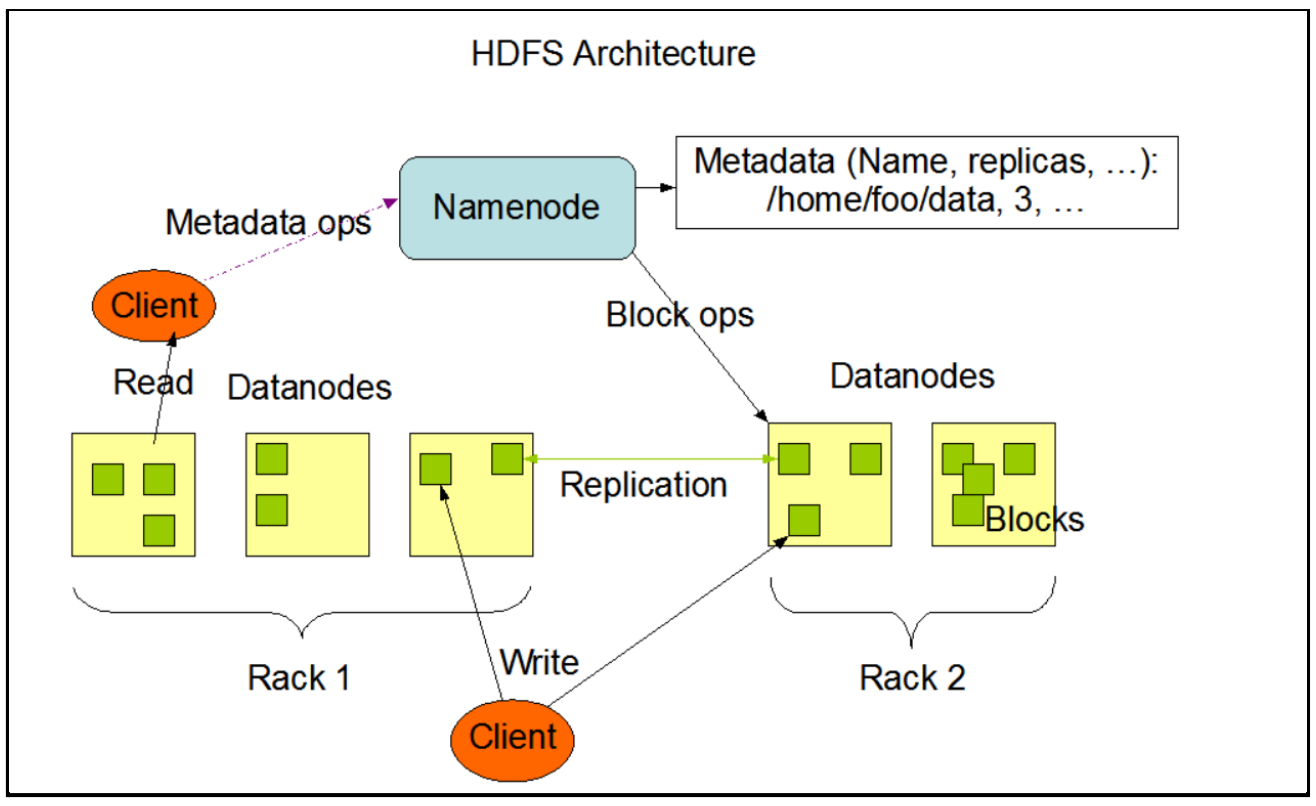

**Fig 9: HDFS Architecture**

Figure 9 shows a HDFS master/slave architecture. An HDFS cluster consists of a single NameNode, a master server that manages the file system namespace and regulates access to files by clients. In addition, there are several DataNodes, usually, one per node in the cluster, which manage storage attached to the nodes that they run on. HDFS exposes a file system namespace and allows user data to be stored in files. The NameNode executes file system namespace operations like opening, closing, and renaming files and directories. It also determines the mapping of blocks to DataNodes. The DataNodes are responsible for serving read and write requests from the file system's clients. The DataNodes also perform block creation, deletion, and replication upon instruction from the NameNode.

### 3.2.1 Data Storage in HDFS

HDFS supports a traditional hierarchical file organization. A user or an application can create directories and store files inside these directories. The file system namespace hierarchy is like most other existing file systems; one can create and remove files, move a file from one directory to another, or rename a file. Internally, a file is split into one or more blocks and these blocks are stored in a set of DataNodes.

The NameNode maintains the file system namespace. Any change to the file system namespace or its properties is recorded by the NameNode. An application can specify the number of replicas of a file that should be maintained by HDFS. The number of copies of a file is called the replication factor of that file. This information is stored by the NameNode.

#### 3.2.1.1 Data Blocks

HDFS is designed to support very large files. Applications that are compatible with HDFS are those that deal with large data sets. These applications write their data only once, but they read it one or more times and require these reads to be satisfied at streaming speeds. HDFS supports write-once-read-many semantics on files. A typical block size used by HDFS is 64 MB. Thus, an HDFS file is chopped up into 64 MB chunks, and if possible, each chunk will reside on a different DataNode.

### 3.2.2 Data-Access in HDFS

All HDFS communication protocols are layered on top of the TCP/IP protocol. A client establishes a connection to a configurable TCP port on the NameNode machine. It talks to the NameNode with the ClientProtocol. The DataNodes talk to the NameNode using the DataNode Protocol. A Remote Procedure Call (RPC) abstraction wraps both the Client Protocol and the DataNode Protocol. By design, the NameNode never initiates any RPCs. Instead, it only responds to RPC requests issued by DataNodes or clients.

### 3.2.3 Data-Consistency in HDFS

The HDFS namespace is stored by the NameNode. The NameNode uses a transaction log called the EditLog to persistently record every change that occurs to file system metadata. For example, creating a new file in HDFS causes the NameNode to insert a record into the EditLog indicating this. Similarly, changing the replication factor of a file causes a new record to be inserted into the EditLog. The NameNode uses a file in its local host OS file system to store the EditLog. The entire file system namespace, including the mapping of blocks to files and file system properties, is stored in a file called the FsImage. The FsImage is stored as a file in the NameNode's local file system too.

### 3.2.4 Fault Tolerance in HDFS

The primary objective of HDFS is to store data reliably even in the presence of failures. The three common types of failures are NameNode failures, DataNode failures, and network partitions.

- Data Disk Failure, Heartbeats and Re-Replication

Each DataNode sends a Heartbeat message to the NameNode periodically. A network partition can cause a subset of DataNodes to lose connectivity with the NameNode. The NameNode detects this condition by the absence of a Heartbeat message. The NameNode marks DataNodes without recent Heartbeats as dead and does not forward any new IO requests to them. Any data that was registered to a dead DataNode is not available to HDFS anymore. DataNode death may cause the replication factor of some blocks to fall below their specified value. The NameNode constantly tracks which blocks need to be replicated and initiates replication whenever necessary. The necessity for re-replication may arise due to many reasons: a DataNode may become unavailable, a replica may become corrupted, a hard disk on a DataNode may fail, or the replication factor of a file may be increased.

- Data Integrity

It is possible that a block of data fetched from a DataNode arrives corrupted. This corruption can occur because of faults in a storage device, network faults, or buggy software. The HDFS client software implements checksum checking on the contents of HDFS files. When a client creates an HDFS file, it computes a checksum of each block of the file and stores these checksums in a separate hidden file in the same HDFS namespace. When a client retrieves file contents it verifies that the data it received from each DataNode matches the checksum stored in the associated checksum file. If not, then the client can opt to retrieve that block from another DataNode that has a replica of that block.

- Metadata Disk Failure

The FsImage and the EditLog are central data structures of HDFS. A corruption of these files can cause the HDFS instance to be non-functional. For this reason, the NameNode can be configured to support maintaining multiple copies of the FsImage and EditLog. Any update to either the FsImage or EditLog causes each of the FsImages and EditLogs to get updated synchronously. This synchronous updating of multiple copies of the FsImage and EditLog may degrade the rate of namespace transactions per second that a NameNode can support. However, this degradation is acceptable because even though HDFS applications are very data-intensive in nature, they are not metadata intensive. When a NameNode restarts, it selects the latest consistent FsImage and EditLog to use.

The NameNode machine is a single point of failure for an HDFS cluster. If the NameNode machine fails, manual intervention is necessary. Currently, automatic restart and failover of the NameNode software to another machine is not supported.

### 3.2.5 Recovery

Before a client can write an HDFS file, it must obtain a lease, which is essentially a lock. This ensures the single-writer semantics. The lease must be renewed within a predefined period if the client wishes to keep writing. If a lease is not explicitly renewed or the client holding it dies, then it will expire. When this happens, HDFS will close the file and release the lease on behalf of the client so that other clients can write to the file. This process is called **lease recovery**.

If the last block of the file being written is not propagated to all DataNodes in the pipeline, then the amount of data written to different nodes may be different when lease recovery happens. Before lease recovery causes the file to be closed, it's necessary to ensure that all replicas of the last block have the same length; this process is known as **block recovery**. Block recovery is only triggered during the lease recovery process, and lease recovery only triggers block recovery on the last block of a file if that block is not in the COMPLETE state.

During write pipeline operations, some DataNodes in the pipeline may fail. When this happens, the underlying write operations can't just fail. Instead, HDFS will try to recover from the error to allow the pipeline to keep going and the client to continue to write to the file. The mechanism to recover from the pipeline error is called **pipeline recovery**.

## 4. Comparative Results

The results for Table 2 and Fig 10 help in conforming our selection of Lustre as one of the prominent file system choices in HPC ecosystem. Many supercomputers are using it and the benchmark scores are high compared to Ceph file system. The stable production release date from Table 3 satisfy that Ceph is a recent file system and has newer strategies to solve data consistency, fault-tolerance and recovery problems.

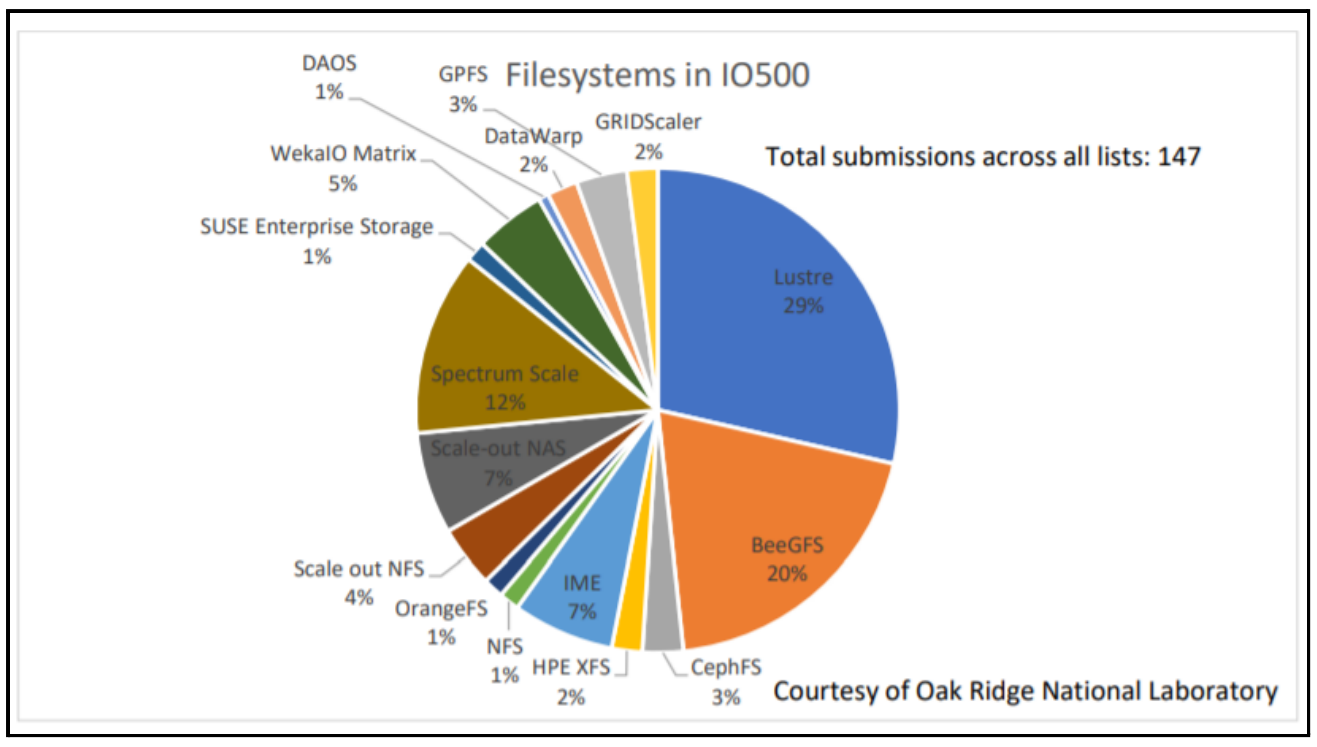

**Fig 10: File systems in IO 500 Supercomputing 2019 edition [37]**

The figure 10 for IO 500 Supercomputing 2019 edition provides insights on the most widely used filesystems in HPC. Lustre filesystem is used by 29% of the HPC system, BeeGFS is the second most widely used filesystem and only 3% of systems use CephFS.

| System | FS | Nodes | client total procs | IOR in GiB/s | MDTest in kIOP/s | Score |
|---|---|---|---|---|---|---|
| Tianhe-2E | Lustre | 480 | 5280 | 209.43 | 982.78 | 453.68 |
| DGX-2H SuperPOD | Lustre | 10 | 400 | 86.97 | 715.76 | 249.5 |
| EC2-10xi3en.metal | CephFS | 10 | 320 | 26.29 | 124.3 | 57.17 |
| TigerShark | CephFS | 15 | 120 | 5.74 | 38.72 | 14.91 |
| Google | Lustre | 1000 | 5000 | 282.78 | 1,148.90 | 569.99 |

**Table 2. Performance of different file systems on the IO-500 list (2021) [35]**

The Table 2 shows results for IO 500 supercomputing 2021 edition. IOR is a parallel IO benchmark that can be used to test the performance of parallel storage systems using various interfaces and access patterns. MDtest is an MPI-based application for evaluating the metadata performance of a file system and has been designed to test parallel file systems. The IOR and MDTest values are the geometric mean of all results. The IO500 score draws an arbitrary equivalency between the difficulty of achieving one gigabyte per second and one kilo-I/O operation per second of performance. The official IO-500 score is computed by sqrt(md * io) with the unit sqrt(GiB*IOP)/s. As we observe for a Lustre file system running on Google system with 1000 Nodes and 5000 processes has a score of 569.99 we can say that it needs 282.78 GiB/s of bandwidth to be equivalent to 1,148.90 KIOP/s. The better way to look at scores is to select them individually based on workload needs since ior-easy and mdtest benchmarks are tests of system capability they demonstrate the peak capability of a system using idealized patterns those real applications strive to generate. On the other hand, the ior-hard benchmark tests an arbitrary pattern that is neither representative of system capability nor any specific user application.

| | Lustre | Ceph |
|---|---|---|
| Type | Parallel File System | Distributed File System |
| Availability | Open Source, Commercial Edition | Open Source, Commercial Edition |
| Founded By | Peter J. Braam in 2001 and had first production release in 2003 | Sage Weil in 2004 and had first stable release in 2012 |
| Who is using? | Sun Microsystems, Silicon Graphics International Corp, Dell, Intel, etc | DigitalOcean, Trendyol Group, Runtastic, SendGrid, etc |
| Written in | C | C++, Python |
| Data-Access | Lustre Network API and Lustre Network driver,<br>LNet layer is connectionless, asynchronous, and does not verify that data has been transmitted while the LND layer is connection-oriented and typically does verify data transmission. | It enables Ceph Clients to directly contact Ceph OSD Daemons. Ceph increases both performance and total system capacity simultaneously, while removing a single point of failure. Ceph Clients can maintain a session when they need to and with a particular Ceph OSD Daemon instead of a centralized server. |
| Usage in Supercomputers (Based on IO500 SC'21 ) | Majority Supercomputers use Lustre filesystem | Few Supercomputers use Ceph Filesystem |
| Website | https://www.lustre.org/ | https://ceph.io/en/ |
| Storage-type | Object Storage | File Storage, Object Storage, Block Storage |
| User-access controls | Access control lists (ACLs) : a list of permissions associated with a system resource | Role Based Access Control (RBACs) : an approach to restricting system access to authorized users. |
| Scalability | Horizontal | Horizontal |
| Partitioning methods | Data Striping - stripe data across multiple OSTs in a round-robin fashion | The Ceph storage system supports the notion of 'Pools', which are logical partitions for storing objects |
| Failures | Client (compute node) failure<br>MDS failure (and failover)<br>OST failure (and failover) | Client failure;<br>MDS failure;<br>OSD failure; |
| Consistency | - It has few consistency issues like Dangling reference, Orphan Objects and Repeated reference. Following are the solutions provided by framework,<br>- **FID-in-LMA:** Lustre* object stores its FID in the XATTR_NAME_LMA extended attribute (EA) for related OI mapping consistency self-verification.<br>- **linkEA:** The MDT-object stores its position (in namespace) information (the name and the parent FID) as XATTR_NAME_LINK EA.<br>- **parent FID for OST-object:** The OST-object stores the FID of its parent MDT-object that references the OST-object as XATTR_NAME_FID EA.<br>- To verify consistency it provides Lustre* consistency verification tools - **LFSCK** that can verify the objects in the whole/partial system | - **Data Scrubbing:** Ceph OSD Daemons can compare their local objects metadata with its replicas stored on other OSDs which is called scrubbing.<br>- Scrubbing happens on a per-placement group base. Scrubbing (usually performed daily) catches mismatches in size and other metadata.<br>- Ceph OSD Daemons also perform deeper scrubbing by comparing data in objects bit-for-bit with their checksums.<br>- Deep scrubbing (usually performed weekly) finds bad sectors on a drive that wasn't apparent in a light scrub. |
| Fault Tolerance | There are two types of failover configurations available: active/passive pair and active/active pair.<br>- **Active/passive pair** - In this configuration, the active node provides resources and serves data, while the passive node is usually standing by idle. If the active node fails, the passive node takes over and becomes active.<br>- **Active/active pair** - In this configuration, both nodes are active, each providing a subset of resources. | - If an MDS daemon stops communicating with the monitor, the monitor will wait mds_beacon_grace seconds (default 15 seconds) before marking the daemon as laggy.<br>- If the standby is available, the monitor will immediately replace the laggy daemon. Each file system may specify the number of standby daemons to be considered healthy.<br>- This number includes daemons in standby-replay waiting for a rank to fail. The pool of standby daemons not in replay count towards any file system count. |
| Recovery | - "**Imperative Recovery**" feature allows the MGS to actively inform clients when a target restarts after a failure, failover, or other interruption to speed up recovery.<br>- "**Metadata Replay**" provides information on recovering from a corrupt file system.<br>- **"Commit on Share"** provides information on resolving orphaned objects, a common issue after recovery<br>- **Snaphshots and Backups** | - **Metadata damage and repair**: If a file system has inconsistent or missing metadata, it is considered damaged. You may find out about damage from a health message, or in some unfortunate cases from an assertion in a running MDS daemon. Metadata damage can result either from data loss in the underlying RADOS layer (e.g. multiple disk failures that lose all copies of a PG) or from software bugs. CephFS includes some tools that may be able to recover a damaged file system, but to use them safely requires a solid understanding of CephFS internals.<br><br>- **Data pool damage** (files affected by lost data PGS): If a PG is lost in a data pool, then the file system will continue to operate normally, but some parts of some files will simply be missing (reads will return zeros). Files are split into many objects, so identifying which files are affected by the loss of particular PGs requires a full scan overall object IDs that may exist within the size of a file. This type of scan may be useful for identifying which files require restoring from a backup. |

**Table 3 Qualitative overview of the HPC storage systems Lustre and Ceph.**

Table 3 presents a qualitative comparison between Lustre and Ceph file systems. It covers various generic aspects like programming language used, stable production version release date, website, data-storage, data-consistency, etc. The basic building block for the storage in both systems is object storage. Being a parallel filesystem Lustre uses data-striping to stripe data across multiple OSTs. Ceph is a distributed file system and uses CRUSH algorithm to decide data storage. Ceph uses a subset of Lustre fault tolerance mechanisms which is active/passive strategy. Lustre offer various types of recovery mechanism over Ceph. The placement strategy used by CRUSH helps in maintaining data-consistency with data-scrubbing mechanisms while in Lustre it requires more tools and methods to achieve data-consistency.

| | Ceph | Lustre |
|---|---|---|
| **ARCHITECTURAL FEATURES**<br>(These features characterize a file system's design and describe the vision of principal architectural approaches that to define a file system's components and essence of internal interactions between of its) | (1) Systems at the petabyte scale are inherently dynamic;<br>(2) Decouple data and metadata operations;<br>(3) Adaptive distributed metadata cluster architecture;<br>(4) Clients;<br>(5) Near-POSIX file system interface;<br>(6) Cluster of OSDs (Object Storage Devices);<br>(7) Metadata servers cluster;<br>(8) Dynamic distributed metadata management;<br>(9) Reliable Autonomic Distributed Object Store (RADOS);<br>(10) Extent and B-tree based Object File System (EBOFS). | (1) Cluster;<br>(2) Management server (MGS);<br>(3) Object-based filesystem;<br>(4) Metadata servers (MDSs);<br>(5) Object storage servers (OSSs);<br>(6) Clients;<br>(7) Object Storage Target (OST);<br>(8) Standard POSIX I/O system calls;<br>(9) Metadata Target (MDT);<br>(10) RAID 0 pattern (data is "striped" across a certain number of objects). |
| **PERFORMANCE FEATURES**<br>(These features characterize a file system's internal techniques that are used for achieving high performance) | (1) Controlled Replication Under Scalable Hashing (CRUSH);<br>(2) Dynamic hierarchical partition;<br>(3) Optimization for the most common metadata access scenarios;<br>(4) No file allocation metadata is necessary;<br>(5) In-memory cache;<br>(6) Lazily flushed journals strategy;<br>(7) Inodes embedded in directory;<br>(8) Partitioning the directory hierarchy across multiple nodes;<br>(9) Knowledge of metadata popularity;<br>(10) Hashing content of large or heavy load directories by file name across the cluster;<br>(11) Specially optimized low-level disk scheduler;<br>(12) B-tree service of EBOFS. | (1) Modified version of the ext4 journaling file system;<br>(2) Lustre supports mmap() file I/O;<br>(3) Lustre uses RAID-0 striping and balances space usage across OSTs;<br>(4) MPI ADIO layer that optimizes parallel I/O;<br>(5) Ability to stripe data across multiple OSTs in a round-robin fashion. |
| **SYNCHRONIZATION FEATURES**<br>(Synchronization refers to one of two distinct but related concepts: synchronization of processes, and synchronization of data. Synchronization features provide file system's internal techniques and architectural primitives being used to implement data synchronization) | (1) Object locks;<br>(2) O_LAZY flag (allows applications to explicitly relax the usual coherency requirements for a shared-write file);<br>(3) No metadata locks or leases are issued to clients;<br>(4) Capabilities (specifying which operations are permitted);<br>(5) Shared long-term storage and carefully constructed namespace locks. | (1) In a cluster most operations are atomic;<br>(2) Distributed lock manager (DLM);<br>(3) Two types of request: lock related and data related. |
| **RELIABILITY FEATURES**<br>(Reliability features are internal techniques and approaches that make possible to oppose against unfavourable factors (for example, Sudden Power-Off) and to keep data in consistent state) | (1) Commit metadata updates to disk;<br>(2) Lazily flushed journals of MDS;<br>(3) Quick rescan of MDS journal by any node in the case of MDS failure;<br>(4) OSDs self-report;<br>(5) Active monitoring of OSDs peers in PG;<br>(6) OSD liveness (OSD reachable + assigning data by CRUSH);<br>(7) Object version number + PG's log of recent changes;<br>(8) Fast Recovery Mechanism (FaRM). | (1) Checksum of all data sent from the client to the OSS;<br>(2) Distributed file system check (lfsck) |
| **HIGH-AVAILABILITY FEATURES**<br>(High-availability features provide the ability of the user community to access the system, whether to submit new work, update or alter existing work, or collect the results of previous work) | (1) Cluster of OSDs;<br>(2) Uniform striping and distribution strategy;<br>(3) Placement Groups (PG) + Controlled Replication Under Scalable Hashing (CRUSH);<br>(4) Failure is the norm;<br>(5) Primary-copy replication;<br>(6) OSD monitor | (1) Active/active failover using shared storage partitions for OSS targets (OSTs);<br>(2) Active/passive failover using a shared storage partition for the MDS target (MDT);<br>(3) High availability (HA) manager;<br>(4) Multiple mount protection (MMP) provides integrated protection from errors;<br>(5) Availability is accomplished by replicating hardware and/or software;<br>(6) A pair of servers with a shared resource |
| **NAMESPACE FEATURES**<br>(These features represents special approaches and techniques that to make possible to represent data by means of hierarchy of files or in any other way.) | (1) CRUSH;<br>(2) Dynamic Subtree Partitioning;<br>(3) MDS cluster;<br>(4) Object names simply combine the file inode number and the stripe number;<br>(5) Directory's content distribution strategy;<br>(6) Ranges of inode numbers;<br>(7) Auxiliary anchor table (rare inode with multiple hard links);<br>(8) Single authoritative MDS;<br>(9) Popularity of metadata;<br>(10) Three groups of inode contents with different consistency semantics (security, file, and immutable). | (1) Single, coherent, synchronized namespace;<br>(2) POSIX-compliant filesystem;<br>(3) Extended attribute (EA) describes the mapping between file object id and its corresponding OSTs;<br>(4) Each filename points to an inode. The inode contains all of the file attributes. |
| **SECURITY FEATURES**<br>(Security features characterize file system's approaches and techniques that to protect against data corruption or lost because of any malicious actions.) | (1) Capabilities (specifying which operations are permitted) | (1) TCP connections only from privileged ports;<br>(2) Group membership handling is server-based;<br>(3) ACLs |
| **SCALABILITY FEATURES**<br>(Scalability is the ability of a system, network, or process, to handle a growing amount of work in a capable manner or its ability to be enlarged to accommodate that growth.) | (1) Object-based storages;<br>(2) Object names are constructed using the inode number, and distributed to OSDs using CRUSH;<br>(3) MDS response content;<br>(4) Future metadata operations are directed at the authority (for updates) or a random replica (for reads) based on the deepest known prefix of a given path | (1) A new OSS with OSTs can be added to the cluster without interrupting any operations |
| **NETWORK FEATURES**<br>(This class of features describes architectural solutions that to make file system services available by means of using different networkoriented technologies.) | - | (1) Remote Direct Memory Access (RDMA) for Infiniband (OFED);<br>(2) Re-exported using NFS or CIFS (via Samba);<br>(3) All client/server communications in Lustre are coded as an RPC request and response;<br>(4) Lustre Networking (LNET). |

**Table 4 Feature classification for Ceph and Lustre filesystems [38].**

Table 4 presents the comparison between Lustre and Ceph file systems using broad feature classification. The features like

- Performance: To characterize a file system's internal techniques that are used for achieving high performance.
- Reliability: The internal techniques and approaches that make possible to oppose against unfavourable factors (for example, Sudden Power-Off) and to keep data in consistent state.
- Security: To characterize file system's approaches and techniques that to protect against data corruption or lost because of any malicious actions.
- Scalability: The ability of a system, network, or process, to handle a growing amount of work in a capable manner or its ability to be enlarged to accommodate that growth.
- Architectural: To characterize a file system's design and describe the vision of principal architectural approaches that to define a file system's components and essence of internal interactions.
- Synchronization: To provide file system's internal techniques and architectural primitives being used to implement data synchronization.
- High-availability: To provide the ability of the user community to access the system, whether to submit new work, update or alter existing work, or collect the results of previous work.

- Namespace: This feature represents special approaches and techniques to make possible to represent data by means of hierarchy of files or in any other way.
- Network: To describe architectural solutions to make file system services available by means of using different network-oriented technologies

Based on feature listing we can see that Ceph provides more strategies for namespace, reliability, and scalability management. Lustre provides better availability, network, and performance features.

| Filesystem | Design Goals | Top 3 Features | Case Studies |
|---|---|---|---|
| Ceph | (1) To be scalable file system;<br>(2) To utilize a highly adaptive distributed metadata cluster architecture;<br>(3) To be a highly-available file system | 1. **Namespace features** (These features represents special approaches and techniques to represent data by means of hierarchy of files or in any other way)<br>- CRUSH: (Controlled, Scalable, Decentralized Placement of Replicated Data) Algorithm uses intelligent data replication to ensure resiliency, which is better suited to hyper-scale storage.<br>- Single authoritative MDS and MDS cluster;<br>- Object names simply combine the file inode number and the stripe number;<br>- Directory's content distribution strategy;<br><br>2. **Reliability features**<br>- Commit metadata updates to disk;<br>- Quick rescan of MDS journal by any node in the case of MDS failure;<br>- OSDs self-report;<br>- Active monitoring of OSDs peers in PG;<br><br>3. **Performance optimization features**<br>- Dynamic hierarchical partition: Partitioning the directory hierarchy across multiple nodes;<br>- Optimization for the most common metadata access scenarios : using In-memory cache(a data storage layer that sits between applications and databases to deliver responses with high speeds by storing data from earlier requests or copied directly from databases);<br>- Inodes embedded in directory; Inodes are a data structure in a Unix-style file system that describes a file-system object such as a file or a directory;<br>- Hashing content of large or heavy load directories by file name across the cluster;<br>- B-tree service of EBOFS. (an extent and B+tree based object file system, allows arbitrarily sized objects and preserves intra-object locality of reference by allocating data contiguously on disk) | 1. US Signal scales its Storage As A Service (STaaS) offering to help meet growing demand<br>2. Capturing flight data at the edge, for quick processing in geographically dispersed data centers<br>3. THG Ingenuity's mission: to build a state of the art platform that not only supports, but accelerates their ecommerce business |
| Lustre | (1) To be scalable for use in large computer clusters;<br>(2) To support a variety of high performance, low latency networks;<br>(3) To tune the file I/O to specific application requirements;<br>(4) To optimize parallel I/O. | 1. **High-availability features**<br>- Active/active failover using shared storage partitions for OSS targets (OSTs);<br>- Active/passive failover using a shared storage partition for the MDS target (MDT);<br>- High availability (HA) manager;<br>- Multiple mount protection (MMP) provides integrated protection from errors;<br>- Availability is accomplished by replicating hardware and/or software;<br><br>2. **Performance optimization features**<br>- Modified version of the ext4 journaling file system;<br>- Lustre supports mmap() file I/O : multiple processes accessing data in a read only fashion from the same file;<br>- Lustre uses RAID-0 striping and balances space usage across OSTs;<br>- MPI ADIO(Abstract-Device Interface for I/O) layer that optimizes parallel I/O;<br>- Ability to stripe data across multiple OSTs in a round-robin fashion.<br><br>3. **Network features**<br>- Remote Direct Memory Access (RDMA) for Infiniband (OFED);<br>- Re-exported using NFS or CIFS (via Samba);<br>- All client/server communications in Lustre are coded as an RPC request and response;<br>- LNet and LND: LNet layer is connectionless, asynchronous, and does not verify that data has been transmitted while the LND layer is connection-oriented and typically does verify data transmission. | 1. Maxar Uses Amazon FSx for Lustre to Deliver Forecasts Weather<br>2. Toyota Research Institute (TRI) collects and processes large amounts of sensor data from their autonomous vehicles (AV) test drives using Amazon FSx for Lustre.<br>3. Netflix VFX team uses Amazon FSx for Lustre. |

**Table 5 Design Goals vs Key Features of Lustre and Ceph File Systems [38]**.

Table 5 maps key design goals and features of Lustre and Ceph file systems. The case-studies of various companies using it, helps for making a logical decision while picking a storage system with similar design goal or feature aspects in HPC ecosystem. Ceph is suitable for high-latency applications, while Lustre is suitable for low-latency applications.

| | CockroachDB | HDFS |
|---|---|---|
| Type | NewSQL Storage Engine (Distributed FS) | Distributed File System |
| Availability | Open Source, Community Edition, Commercial Edition | Open Source, Commercial Edition |
| Developed By | Spencer Kimball at cockroach labs in 2014 | Doug Cutting at Apache Software Foundation in 2006 |
| Who is using? | DoorDash, DevSisters, MUX, etc | Uber, Airbnb, Netflix, etc |
| Written in | Go | Java |
| Primary database model | Relational DBMS | Wide column store in HBase;<br>Relational DBMS in Hive; |
| DB-engines Ranking | 58 / 353 | 25 / 353 HBase;<br>14 / 353 Hive; |
| Website | www.cockroachlabs.com | hbase.apache.org<br>hive.apache.org |
| Storage-type | Key-Value Pairs | Blocks |
| User-access controls | Role-based access control | Access Control Lists (ACL) for RBAC, integration with Apache Ranger for RBAC & ABAC |
| Scalability | Horizontal, automated scaling | Horizontal |
| Consistency | full support for ACID transaction semantics using Parallel Commits | one-copy-update-semantics -all clients see contents of file identically as if only one copy of file existed. |
| Replication methods | horizontal partitioning (by key range) | Sharding |
| Partitioning methods | Multi-source replication using RAFT | Multi-source replication<br>Source-replica replication |
| MapReduce Support | no | yes |
| Foreign keys support | yes | no |
| Failures | Offline Nodes; | NameNode failures, DataNode failures, and network partitions |
| Fault Tolerance | Replication using RAFT protocol - It is a consensus protocol—an algorithm that makes sure that your data is safely stored on multiple machines, and that those machines agree on the current state even if some of them are temporarily disconnected.<br>Rebalancing Nodes; | Heartbeats and Re-Replication;<br>Data Integrity - To verify data corruption perfrom checksum<br>Copy of FSImage - Incase NameNode fails, load copy of FSImage on new NameNode |
| Recovery | Snapshots | **lease recovery** - used,If a node having lock/lease for a file dies;<br>**block recovery** - used, if the last block of the file being written is not propagated to all DataNodes in the pipeline;<br>**pipeline recovery** - used, if performing write pipeline operations, some DataNodes in the pipeline fail.; |

**Table 6 Qualitative comparison of CockroachDB and HDFS.**

| DB Engines | Design Goals | Key Features | Applications |
|---|---|---|---|
| CockroachDB | to support for ACID transactions;<br>to have horizontal scalability;<br>to have strong consistency;<br>to have survivability i.e. to have minimal latency disruption and no manual intervention with disk, machine, rack, and even datacenter failures. | - **EFFORTLESS SCALE**: Scale your applications, not operational complexity. CockroachDB scales without manual operation, eliminates overhead associated with manual sharding, and balances server utilization to reduce costs.<br>- **BULLETPROOF RESILIENCE**: Ensure your apps are always on and available. CockroachDB survives system failures without interruption, reduces the need for complex and costly backup configurations, and eliminates the risk of losing data.<br>- **ENSURED CONSISTENCY**: Guarantee transactional consistency in distributed SQL. CockroachDB ensures all your data is always correct and up-to-date, guarantees serializable isolation, and applies a cost-based optimizer and query planner to distributed workloads.<br>- **GEO-LOCALITY:** Give your data a location. CockroachDB is the only database that assigns data to a location allowing you to reduce latencies, move data close to your customers, and remain globally compliant. | System of record (financial ledger, inventory management, transaction record),<br>identity access management (IAM),<br>metadata layer,<br>general purpose DB |
| HDFS | to store and process large amounts of data;<br>to have random access storage and retrieval data platform;<br>to store variety of data storage formats;<br>to use map and reduce;<br>to have auto failover and reliability; | - No single point of failure ensures very **high availability** with multiple customers having achieved 100% availability on a 3+ year lookback basis.<br>- **Operational simplicity** for lowest total cost of ownership.<br>- Best-in-class **scalability and performance** of NoSQL platforms.  No need for managing and redesigning shards as the cluster grows.<br>- Additional modalities with **add-on applications**: Apache Phoenix enables a SQL interface, OpenTSdb enables Time series operations, JanusGraph enables a scale-out graph interface and GeoMesa provides a geo-spatial interface. | Internet of Things (IOT),<br>fraud detection applications,<br>recommendation engines,<br>product catalogs and playlists,<br>messaging applications,<br>customer 360 applicaitions,<br>web applications,<br>machine learning model serving. |

**Table 7 Design Goal vs Key Features of HDFS and CockroachDB. [38]**

Table 6 presents qualitative comparison of HDFS and CockroachDB. The DB Engine scores clearly convey why HDFS is one of the prominent choices in BDA ecosystem. The production release for both the storage systems conform that CockroachDB is a newer system and uses NewSQL techniques. These techniques need more time to have stable build as they are still in early phase. The basic storage unit of both the systems are key-value for CockroachDB and blocks for HDFS. CockroachDB support automated scaling which is a big relief over manual scaling. With the use of Parallel commits, CockroachDB achieves its ACID properties in distributed environment, while HDFS simply uses one-copy-update semantics. Recovery mechanism provided by HDFS are more sophisticated.

Table 7 maps key design goals and features of HDFS and CockroachDB. The variety of use-case mapping these systems helps for making a logical decision while picking a storage system with similar design goal or feature in a BDA ecosystem. CockroachDB is well-suited for systems requiring ACID compliance, strong consistency, and minimal downtime. HDFS is well-suited for storing large-amount of data and variety of data-formats, strong fault-tolerance and recovery needs.

## 5. Conclusion

Storage systems are a combination of both hardware storage devices and software file systems which serves the purpose of all storage related issues. There are various applications of HPC and BDA ecosystems like Modular Ocean Model, mpiBLAST, ECOlogical model, social media, Sensor Data, etc. Such applications can be broken down into smaller subsets of problems. The design goals can be evaluated for those problems and by aggregating and comparing those design goals with features provided by available file system a logical decision can be made on selecting an existing filesystem or creating a newer filesystem based on the needs of the application. In our paper we have mapped design goals of file systems to various key features and have supported them with use-cases being used in industry.

Based on the qualitative comparison we can comment that the object-based storage and distributed file systems are basic similarities among the filesystems today. For HPC ecosystem still Lustre is one of the most prominent choices as per IO500 SC'21 results [27]. Hadoop (HDFS) is prominent choice for many Big-Data applications as per DB Engines Ranking [39].

Ceph still has some drawbacks. Among them is the limitation of only being able to deploy one CephFS per cluster and the current test phase of reliability on real-world use-cases. Some features and utilities are still in an experimental phase as well. For instance, usage of snapshots could cause client nodes or MDSs to terminate unexpectedly. In addition, Ceph is designed with HDDs as its basis and needs improvements in performance when disks are replaced with SSDs, and the data access pattern is random [23]. HDFS based storage solutions will provide low latency over cloud deployed CockroachDB solution.

We have compared four storage systems where in two storage system from each environment. There are hundreds of such file system which need similar assessment. Finally, it comes down to the storage system architect to understand the design goals of the application for his problem and to map it with the best available features and solution in the market with proven performance.

## 6. Future Developments

On the hardware side, NVRAM storage will likely transform how we build storage systems. On the one hand, NVRAM can improve the capabilities to record large amounts of data at the pace required to be useful for later analysis tasks. On the other hand, NVRAM can dramatically simplify storage systems, which currently add complexity to every effort for relatively modest performance improvements. [2]

There is a wide variety of new and upcoming approaches for file and storage systems. Their motive is optimization and improvement is highly required due to the challenges regarding managing the vast amount of data from I/O-intensive applications. The HPC community aims to relax the strict POSIX semantics without losing the support for legacy applications. New approaches like DAOS: It will support performing I/O in a scalable way, so that multiple processes can perform asynchronous write operations without having to worry about consistency problems [35]. Týr is a blob storage system with support for transactions; it provides blob storage functionality and high access parallelism, which is enough for converging some applications in HPC and BDA like BLAST, MOM, ECO-HAM and Ray Tracing applications [1], SoMeta: Scalable Object-Centric Metadata Management (SoMeta) is intended for future object-centric storage systems, providing the corresponding metadata infrastructure. A distributed hash table is used to organize metadata objects that contain the file system metadata [36].

Storing bigdata in a decentralized storage is one of the solutions for ever-growing data. In order to mitigate the challenges like scalability and transaction speeds in decentralized storage, techniques like swarming [38] and sharding [37] are being used. Partitioning of databases along logical lines is referred to as sharding. The decentralized model ensures the storage of shards together. Moreover, a unique partition key is used by a dedicated decentralized application to access the shards. Besides this, swarming is used to enable the collective storage of shards. Data is stored and managed by creating a large group of nodes, which is called a swarm. This group of nodes is similar to the network of nodes created for blockchain. Possibly decentralized storage can be a global solution later in the future for big data storage systems. [39]